\documentclass[aps,prd,reprint,showkeys,showpacs,nofootinbib]{revtex4-1}

\usepackage{bm}
\usepackage{graphicx} %
\usepackage{amsmath} %
\usepackage{amssymb} %
\usepackage{url}
\usepackage{subfigure} %
\usepackage{float} %
\usepackage{upgreek} %
\usepackage{multirow} %
\usepackage{color} %
\usepackage{lineno} %
\usepackage{hyperref} %
\usepackage{booktabs}
\usepackage{placeins}
\usepackage[usenames,dvipsnames]{xcolor}
\hypersetup{colorlinks=true, urlcolor=black, linkcolor=blue, citecolor=red} %

\begin{document}

\title{Test of the Einstein equivalence principle with spectral distortions in the \\ cosmic microwave background}

\author {Shun Arai${}^1$}
\email{arai.shun@a.mbox.nagoya-u.ac.jp}

\author {Daisuke Nitta${}^1$}

\author {Hiroyuki Tashiro${}^1$}
\affiliation {${}^1$Department of Physics and Astrophysics, Nagoya University, Nagoya 464-8602, Japan}
\date{\today}

\begin{abstract}
 
The Einstein equivalence principle (EEP) can be verified by the
measurement of the spectral distortions of the cosmic microwave background (CMB).  One of the consequences of the EEP on cosmological
scales is the energy independency of the cosmological redshift effect.
We propose a new test of the energy independency of the redshift effect
by the measurement of the spectral distortion of CMB.  In general relativity, the energy independency of the
redshift effect is ensured by the Friedmann-Robertson-Walker (FRW) metric which does not depend on energy. 
We show that the CMB spectral distortions arise when the FRW metric has
the energy dependence. Assuming the simple energy-dependent form of the
FRW metric, we evaluate the CMB distortions.
From the COBE/FIRAS bound, we find that the deviation degree from the
 EEP is, at least, less than $10^{-5}$ at the CMB energy scales.


\end{abstract}

\maketitle

\section{\label{sec:level1}Introduction}

Observations of the cosmic microwave background (CMB) have become essential
tools in modern cosmology. Precise measurements of the CMB temperature
and polarization anisotropies provide valuable information about the
Universe~\cite{2015arXiv150201582P}. Recently, the measurement of CMB spectral distortions, that
is, the deviation of the CMB frequency spectrum from a blackbody
spectrum, has been expected as a new cosmological probe.

COBE/FIRAS has obtained the almost perfect blackbody spectrum of the CMB
with the temperature $T_0 = 2.726$~K~\cite{2009ApJ...707..916F}.
Although they have not been detected yet,
CMB distortions can be generated within the
standard cosmological model
as well as with new physics~(for reviews, see
\cite{2012MNRAS.419.1294C,2013IJMPD..2230014S,2014PTEP.2014fB107T}).
Currently, the observational bound on the spectral distortions is given in terms of
two types of distortions, $\mu$ and $y$-type distortions~\cite{1969Ap&SS...4..301Z,1970Ap&SS...7...20S}.
The $\mu$-type distortion is described with a nonvanishing chemical
potential~$\mu$ and created at $10^6 \gtrsim z \gtrsim 5 \times 10^4$
where,
even if the CMB spectral distortionsarise,
Compton
scattering is efficient enough to maintain the kinetic equilibrium of
CMB photons.
The $y$-type distortion is parametrized by the Compton $y$ parameter and
generated in lower redshifts $z<5 \times 10^4$ where, once
the CMB spectrum is distorted, the kinetic
equilibrium of CMB photons is no longer maintained.
The current constraints on the distortion parameters are provided by
COBE/FIRAS as $|y| < 1.5\times10^{-5}$ and $|\mu|<9\times 10^{-5}$~\cite{1996ApJ...473..576F}.  
To improve these bounds, next-generation CMB spectrometers are being
discussed~\cite{2011JCAP...07..025K,2014JCAP...02..006A}.
The future measurements or constraints on the CMB distortions
allow us to
access
the properties of primordial fluctuations~\cite{primordial},
the nature of dark matter~\cite{DM}, the abundance of primordial black
holes~\cite{Tashiro:2008sf}, the existence of primordial magnetic
fields~\cite{Jedamzik:1999bm} and other high-energy
physics~\cite{Tashiro:2012nb}.
In this paper, we discuss that
the measurement of CMB distortions can also test general relativity~(GR), in particular,
the Einstein equivalence principle~(EEP).

Since GR was proposed as the theory of
gravity by Einstein, the theory has passed almost all tests
such as ground-based and Solar System experiments~\cite{2006LRR.....9....3W}. And
furthermore, the gravitational wave detection by LIGO proves the
accuracy of the theory even in a strong gravitational field~\cite{LIGO2, LIGO}.
However, it still leaves room to verify GR at the cosmological scales.
Since the first evidence was presented by
the type-Ia supernova observations~\cite{1998AJ....116.1009R,1999ApJ...517..565P},
independent cosmological observations strongly support the accelerating expansion of
the Universe.
As an origin of this acceleration, GR
requires the existence of unknown dark energy.
Alternatively,
the modification of GR on cosmological scales
is suggested to explain the acceleration 
as an effect of gravity~\cite{2010PhRvD..81j3510Z,2012PhR...513....1C,2016RPPh...79d6902K}.
Therefore, 
it is still quite important to verify GR in the cosmological context,
and
we pay attention to
the validity of the EEP which is
one of the fundamental principles in GR.

The EEP is tested from laboratory to Solar System
scales by many authors~(for reference, see Ref.~\cite{2001CQGra..18.2397A}).
In these studies, the validity of the EEP has been obtained
from the travels of a test particle through the gravitational potential.
Therefore, as the constraint on the EEP, these studies have provided the constraint on
the energy dependency of parametrized-post-Newtonian (PPN) parameter $\gamma$.
Recently, 
this energy dependency has been also tested 
by using high-energy photons emitted by gamma ray bursts, fast radio bursts,
and TeV blazers with the gravitational potential of the Milky Way~\cite{Abdo2009, 1999BASI...27..627S,2015ApJ...810..121G, 2015PhRvL.115z1101W, 2016ApJ...818L...2W,2016ApJ...821L...2N,2016arXiv160104558Z,2016ApJ...820L..31T}.

In this paper,
we focus on 
the independency of the cosmological redshift effect on the energy of a
test particle.
Because this energy independency is one of the consequences of the EEP on cosmological scales,
it is important to test the independency of the
redshift effect by cosmological observations.  We show that
the independency of the redshift effect can be verified by measurement of
the CMB distortion.  After submitting our paper,
Ref.~\cite{Ferreras:2016xsq} appeared.  They have investigated the
energy dependence of the cosmological redshift effect using the emission
lines over the 3700--6800~\AA~range in SDSS spectroscopic data at $0.1 <
z <0.25$.  Their conclusion is that they cannot find any
energy-dependence of the redshift with a precision of $10^{-6}$ at
$z<0.1$ and $10^{-5}$ at $0.1<z<0.25$.  Our method is complementary with
theirs because probing energy is different. Besides, CMB distortion can
verify the EEP up to redshifts larger than $z\sim 1000$.

To demonstrate the test of the EEP through the CMB distortion, we
introduce a simple energy dependence of the
Friedmann-Robertson-Walker~(FRW) metric.  Generally, when a metric
depends on energy, the EEP is violated in this metric theory of gravity.
In other words, the existence of the energy dependence of the metric
means that the structure of spacetime felt by a test particle depends on
its own energy.

In GR, although CMB photons are redshifted due to the
cosmic expansion, the blackbody spectrum of the CMB is hold during their
free-streaming because the EEP ensures that redshift effect is
independent of the photon energy.  However, when the redshift effect
depends on the photon energy, the deviation from the blackbody spectrum
arises even in the free streaming regime.
We evaluate the CMB distortion and
obtain the constraints on the accuracy of the EEP on cosmological
time and length scales through a comparison with the COBE/FIRAS data.

\section{energy-dependent FRW metric}

Since the energy dependency of the metric violates the EEP, 
we first consider the energy-dependent FRW metric.
Taking into account the cosmological principle,
we can be allowed to introduce two energy dependent functions, $f(E)$ and
$g(E)$, in the FRW metric as
\begin{align}
ds^2 = -\frac{dt^2}{f^2(E)}+\frac{a^2(t)}{g^2(E)}\delta_{ij}dx^idx^j\,, \label{metric}
\end{align}
where $E$ denotes the energy of a photon observed by a free-falling
observer in this metric, and 
$f(E)$ and $g(E)$ are arbitrary functions of $E$.
Although, without assuming any certain gravity theory, we determine the
form of Eq.~(\ref{metric}) based on the
cosmological principle,
the same energy dependence is often discussed
in rainbow gravity, which is one of the gravity theories without the EEP~\cite{2004CQGra..21.1725M,2007JCAP...08..017L,2010PhLB..687..103L}.

Since the FRW metric depends on the energy, the redshift effect due to
the cosmological expansion also has energy dependence.
To derive the redshift effect, we consider the geodesic equation for a
photon with energy $E$.
In the metric given by Eq.~(\ref{metric}),
nonvanishing Christoffel symbols are 
\begin{align}
&\Gamma^{0}_{00} = -\frac{\dot{f}}{f}\,,\quad
~\Gamma^{0}_{ij} = \left(
 \frac{f}{g}\right)^2\left(a\dot{a}-a^2\frac{\dot{g}}{g}\right)\delta_{ij}\,,
 \nonumber \\
&\Gamma^{i}_{0j} = \left(\frac{\dot{a}}{a}-\frac{\dot{g}}{g}\right)\delta^i_j\,\label{Chi0j},
\end{align}
where the dot denotes the derivative with respect to time.

Therefore, the geodesic equation provides the modified redshift effect,
\begin{align}
\dot{E} = -
\frac{\dot{a}}{a} \left ( 
 {1 -\frac{d \log g} {d \log E}}  \right)^{-1} E
 \label{Eevo}.
\end{align}
When $f$ and $g$ are constant, the redshift effect is the same as in GR.

\section{CMB distortions due to the energy-dependent redshift effect}

After the epoch of recombination,
the universe becomes transparent for photons and they are free to stream out.
During such a free-streaming regime, 
the evolution of the CMB photon energy
distribution is given by the collisionless Boltzmann equation.
Assuming the homogeneity and isotropy of the Universe, the 
collisionless Boltzmann equation in the metric by Eq.~(\ref{metric}) can be
described as 
\begin{align}
\frac{\partial{n_E}}{\partial{t}} -
\frac{\dot{a}}{a} E
 \left( 1 - \frac{d \log g} {d \log E} \right)^{-1} 
 \frac{\partial{n_E}}{\partial{E}} = 0\,\label{Boltz}.
\end{align}
Although the general solution of Eq.~(\ref{Boltz}) is provided in a
function of the combination value, $a E/g$,
we need the initial condition of the energy distribution to solve Eq.~(\ref{Boltz}).

Well before the epoch of recombination, the time scale of thermal
equilibrium for CMB photons is much shorter than the cosmological time
scale. In this regime, when the deviation from a blackbody spectrum
arises, the deviation is quickly erased and the blackbody spectrum
is maintained.
Therefore, for simplicity, we assume that the energy distribution of the
CMB is a blackbody spectrum,~$[\exp(E/T_{\rm re})-1]^{-1}$, at the epoch of
recombination, where
$T_{\rm re}$ is the temperature at that epoch.
However, as mentioned
above, CMB distortions can be generated below $z \sim 10^6$, which is
well before the epoch of the recombination. During this regime, the evolution of
the CMB distortions is provided by the collisional Boltzmann
equation. We will discuss this issue later.

With this assumption, the solution of Eq.~(\ref{Boltz}) is given by
 \begin{equation}
  n_E=
\frac{1}{
\exp[\eta(E, z) E/T_z] -1},
\label{eq:neq_sol}
 \end{equation}
where $T_z = T_{\rm re} (1+z)/(1+ z_{\rm re})$ with
the redshift for the epoch of recombination $z_{\rm re}$, and $\eta(E,
z)$ is provided by
\begin{equation}
 \eta(E,z) = 
  \frac{g(E_{\rm re}(E,z) )}{g(E)},
\end{equation}
where the function $E_{\rm re}(E,z)$ represents the energy at $z_{\rm re}$ for a photon 
whose energy is redshifted to $E$ at the redshift $z$.
We can obtain $E_{\rm re}(E,z)$ from Eq.~(\ref{Eevo}).
Since various tests support the validity of GR, we assume that $g^{-1}$ can
be approximated in $g^{-1} \approx 1 + h(E)$ with $h(E) \ll 1$. In the
leading order of $h(E)$, the function $\eta(E,z)$ can be expanded in 
\begin{equation}
 \eta(E,z) \approx   
1+h(E) - h \left(
	    \frac{1+ z_{\rm re}}{1+z} E
\right).
\label{eq:app_eta}
\end{equation}

The aim of this paper is to obtain the constraint on $h(E)$ from
the measurement of the CMB distortions.
Here we demonstrate two simple cases of the function $h(E)$. In the first
case, $h(E)$ is a linear function of $E$. In the second case,
$h(E) $ is proportional to $E^{-1}$.

\subsection{ The case with $h(E) \propto E $}

We assume that the form of $h(E)$ is given by
\begin{equation}
 h(E) = \delta_{T_0} E/T_0,
\end{equation}
with $\delta_{T_0} \ll 1 $.
Here the parameter
$\delta_{T_0}$ represents the deviation degree from the EEP at the energy
scale~$T_0$.
The CMB photon energy distribution at the present epoch is given from
Eqs.~(\ref{eq:neq_sol}) and (\ref{eq:app_eta}).
Expanding the photon energy distribution
up to the
linear order of $\delta_{T_0}$, we obtain 
\begin{equation}
  n_E \approx \frac{1}{
\exp(E/T_0) -1 } +
  \frac{\exp(E/T_0)}{ [\exp(E/T_0) -1]^2} 
\left(\frac{E}{T_0} \right)^2
z_{\rm re} \delta_{T_0}.
\label{eq:expand-1}
\end{equation}
The first term represents the blackbody spectrum with $T_0$ and the second term
provides the deviation from the blackbody spectrum.

We define the relative deviation from the blackbody spectrum as
$\Delta_E = (n_E - n_{{\rm BB},E}) /n_{{\rm BB},E}$ where
$n_{{\rm BB},E}$ is the blackbody spectrum with $T_0$.
According to Eq.~(\ref{eq:expand-1})
$\Delta_E$ is given by
\begin{equation}
 \Delta_E =  \frac{\exp(E/T_0)}{ \exp(E/T_0) -1} 
\left(\frac{E}{T_0} \right)^2
z_{\rm re} \delta_{T_0}.
\end{equation}

\begin{figure}[t]
  \begin{center} 
    \includegraphics[clip,width=7.0cm]{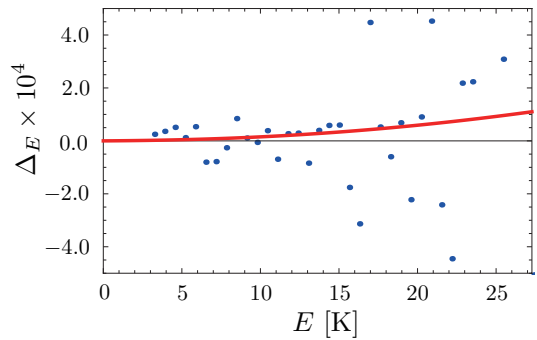}
    \caption{The relative deviation from the blackbody spectrum,
   $\Delta_E$. Here we
   adopt $h(E) = \delta_{T_0 } E/T_0$ with $\delta_{T_0} = 10^{-9}$. The
   $x$ axis is the
   energy of CMB photons in units of Kelvin. Blue points represent the
   residual measured by COBE/FIRAS~\cite{1996ApJ...473..576F}.} 
    \label{fig:const}
  \end{center}
\end{figure}

We show $\Delta_E$ as a function of $E$ in Fig.~\ref{fig:const}.
Here we set $\delta_{T_0}=10^{-9}$ and $z_{\rm re}=1100$.
COBE/FIRAS has provided the
possible residual from the blackbody spectrum~\cite{1996ApJ...473..576F}.
We plot the residual as blue points in Fig.~\ref{fig:const}.
From the figure, we conclude that COBE/FIRAS gives the 
upper bound,
\begin{equation}
|\delta_{T_0}|  \lesssim 10^{-9}. 
\end{equation}

\subsection{ The case with $h(E) \propto E^{-1}$}

Next we consider the case where 
$h(E)$ is represented as
\begin{equation}
 h(E) = \delta_{T_0} T_0/E .
  \label{eq:caseII}
\end{equation}
Similarly to the previous case, we can obtain the CMB photon distribution
from Eqs.~(\ref{eq:neq_sol}) and (\ref{eq:app_eta}).
The CMB photon distribution can
be approximated to 
\begin{equation}
  n_E \approx \left(
\exp\left[\frac{E}{T_0} \left(1+ \frac{z_{\rm re}}{1+z_{\rm re}}
\frac{T_0}{E}\delta_{T_0}    \right )\right] -1 
\right)^{-1}.
\label{eq:expand}
\end{equation}

This corresponds to the Bose-Einstein distribution, 
$n_E = (\exp(E/T_0  + \mu) -1 )^{-1} $, with the dimensionless chemical potential $\mu =
z_{\rm re} \delta_{T_0}/(1+z_{\rm re})$.

COBE/FIRAS provides the constraint on $\mu $ for CMB photons, $|\mu| < 9
\times 10^{-5}$. Therefore we obtain the limit,
\begin{equation}
|\delta_{T_0}| \lesssim 9 \times 10^{-5}. 
\end{equation}
Currently, PIXIE is designed to be 3 orders of magnitude better than
COBE/FIRAS in the sensitivity~\cite{2011JCAP...07..025K}.
The sensitivity of PIXIE is expected to be close to that required to
measure the distortions arising from 
the dissipation of the scale-invariant primordial fluctuations, $\mu \sim 10^{-8}$,
which is one of unavoidable cosmological sources for CMB distortions.
When PIXIE provides the constraint $\mu \lesssim 10^{-8}$, 
the constraint on the EEP reaches
$|\delta_{T_0}|  \lesssim  10^{-7}$ in the case of Eq.~(\ref{eq:caseII}).

\section{Conclusions}

In this paper, we have proposed that the measurement of CMB spectral
distortions can test the accuracy of the EEP on cosmological scales.
The energy independence of the cosmological redshift effect is
one of consequences of the EEP.
When the FRW metric has energy dependence, the EEP is violated on
cosmological scales.
As a result,
the geodesic equation of a photon is modified and 
the redshift effect depends on its energy.
We have shown that,
in the energy-dependent FRW metric,
CMB distortions are generated even in the
free-streaming regime through the energy-dependent redshift effect.
The shape and amplitude of the distortion depends on the form of the energy dependency.

To parametrize the validity of the EEP in the FRW metric,
we have introduced the deviation parameter $\delta_{T_0}$ representing 
the deviation from the EEP
on the CMB energy scale.
We have analytically evaluated
the CMB distortions in two simple power-law cases of the energy-dependent deviation in the FRW
metric with the power law indices $n=1$ and $n=-1$.
In the first case with $n=1$, we have found that the COBE/FIRAS bound indicates that the EEP
is valid within the degree of the deviation, $|\delta_{T_0}| \lesssim 10^{-9}$, on the CMB
energy scale, $0.0001$--$1$~eV.
When $n>0$, the deviations at higher energy scales are larger than at
lower energy scales. This means that, as $n$ becomes larger, the deviation increases at higher redshifts.
Therefore, the constraint on $\delta_{T_0}$ becomes tighter when $n$ increases.
In the second case with $n=-1$, the generated distortion is represented as the $\mu$-type
distortion and the COBE/FIRAS bound provides the constraint 
$|\delta_{T_0}| \lesssim 10^{-5}$.
When $n<0$, the deviations at lower energy scales are larger than at
higher energy scales. Therefore, we obtain
$|\delta_{T_0}| \lesssim 10^{-5}$ for $n<0$.
Depending on the energy dependence of the FRW metric,
the spectral shape is different from the ordinary CMB distortions, $\mu$-
and $y$-type distortions. Therefore, the precise measurement of the
distortion shape can provide us with a strong constraint on the EEP violation.

It is worth summarizing previous works about the test of the EEP and
providing comments on the relevance of our study.
In previous studies, the accuracy of the EEP is investigated with the
energy dependency of the PPN
parameter $\gamma$.
Using the gamma ray observations,
the constraint is provided as $\gamma _{\rm GeV} -\gamma_{\rm
MeV} \lesssim 10^{-8}$ and $\gamma _{\rm eV} -\gamma_{\rm MeV} \lesssim
10^{-7}$~\cite{2015ApJ...810..121G}. In the radio frequency range, the energy difference of
$\gamma$ is less than $10^{-8}$, which is comparable to our results,
from the observations of fast radio bursts~\cite{2015PhRvL.115z1101W,2016ApJ...820L..31T}.
Since the constraints on the energy difference of $\gamma$ is related to
the gravitational potential, these constraints are valid for the
Schwarzschild metric.
Therefore, these constraints cannot be directly applicable to
the FRW metric without taking a theory of gravity.
In Ref.~\cite{Hees:2014lfa}, the authors have discussed that the measurement of CMB distortions can provide the constraint on
the time variation of the fine structure constant due to the EEP violation in the electromagnetic sector.
Our constraint is completely independent of these limits. 
In more detail, we have provided a bound on the EEP in the FRW metric for the cosmological
time scale from the epoch of recombination to the present time.
Additionally, upcoming observations with PIXIE provide 3 orders of magnitude stronger constraints than that of COBE/FIRAS. 
Recently Ref.~\cite{Ferreras:2016xsq} has investigated the energy dependence of the
cosmological redshift with SDSS data.
Their result is consistent with no energy dependence of the redshift
effect 
with a precision of $10^{-6}$ at $z<0.1$ and $10^{-5}$ at $0.1<z<0.25$.
In this work, they used the spectral lines over the 3700--6800~\AA~range
whose energy range is higher than in the CMB observation frequencies.
Although their investigated redshifts are not so high,
their results are complementary with our results. According to both
results, the violation of the EEP in the FRW metric is not found in the
range from microwave to optical frequencies in the current observation precision.

In this paper, we have demonstrated that the measurement of the CMB
distortions can test the EEP in the FRW metric by taking some assumptions.
In particular,
to evaluate the CMB distortions analytically, 
we neglected the evolution of the CMB
distortions before the epoch of recombination.
Although the distortions can be generated in the energy-dependent FRW
metric before this epoch, it is necessary to solve the collisional
Boltzmann equation numerically.
Because the next-generation CMB spectrometers are being planned to
measure CMB distortions precisely, further detailed calculation is required.
We will address these issues for the EEP bound in our future works.

The spectral distortions of the CMB can be generated by other physical
mechanisms, in particular, the processes related to the thermal history
of the Universe.
Therefore, it is difficult to solve these degeneracies to point out the
effect of the EEP violation
by only the CMB distortion measurement.
However, although we have only studied 
the CMB distortions of the CMB,
neutrinos and gravitons also suffer an energy-dependent redshift effect and
their spectra are modified from ones in the standard cosmology
when the EEP is violated in the FRW metric.
Therefore, the frequency spectral measurement of not only CMB but also
neutrinos and gravitational waves can allow us to obtain the
observational suggestion to the EEP violation.


\begin{acknowledgments}
 We thank Naoshi Sugiyama, Yuko Urakawa, and Jens Chluba for the useful discussions.
S.A. and D.N. are in part supported by the Ministry of Education, Culture, Sports, Science and Technology, Japan (MEXT) Grant-in-Aid for Scientific Research on Innovative Areas,~No.~15H05890,
H.T. is supported by Japan Society for the Promotion of Science (JSPS) KAKENHI Grant No.~15K17646 and MEXT's Program for Leading Graduate Schools `` Ph.D. professionals, Gateway to Success in Frontier Asia."
\end{acknowledgments}

\end{document}